\documentclass[9pt,twocolumn,twoside]{osajnl}
\journal{ol} % Choose journal (ao, aop, josaa, josab, ol, pr)
\usepackage{color}
\usepackage{ulem}

\setboolean{shortarticle}{true}
\title{Dynamic coherent perfect absorption in nonlinear metasurfaces}
\author[1,*]{Rasoul Alaee}
\author[1]{Yaswant Vaddi}
\author[1,2]{Robert W. Boyd}
\affil[1]{Department of Physics, University of Ottawa, Ottawa, ON K1N 6N5, Canada}
\affil[2]{The Institute of Optics and Department of Physics and Astronomy, University of Rochester, Rochester NY 14627, USA}
\affil[*]{Corresponding author: rasoul.alaee@gmail.com}

\begin{abstract}
We propose a tunable coherent perfect absorber based on ultrathin nonlinear metasurfaces. The nonlinear metasurface is made of plasmonic nanoantennas coupled to an epsilon-near-zero material with a large optical nonlinearity. The coherent perfect absorption is achieved by controlling the relative phases of the input beams. We show that the optical response of the nonlinear metasurface can be tuned from a complete to a partial absorption by changing the intensity of the pump beam. The proposed nonlinear metasurface can be used to design optically tunable thermal emitters, modulators, and sensors.    
\end{abstract}

\setboolean{displaycopyright}{true}

\begin{document}

\maketitle

\section{Introduction}
An electromagnetic wave carries energy. In many applications in optics (e.g. solar cells, thermal emitters, optical modulators, optical sensors, and optical data processors), it is desirable to absorb light completely~\cite{Salisbury1952,Pozar2009,Watts:2012,Radi2015,Alaee2017}. In principle, complete light absorption occurs when both reflection and transmission vanish at the same frequency. Perfect absorbers have been proposed after the invention of radar by Salisbury, also known as Salisbury screens~\cite{Salisbury1952}. A Salisbury perfect absorber consists of an absorbing layer on top of a metal ground plane separated by a dielectric spacer. Salisbury perfect absorber relies on the destructive interference of the reflected light between the absorbing layer and the metal ground plane (by varying the thickness of the dielectric spacer) and suppression of the transmitted light (due to optically thick metal ground plane).

It is well-known that an optically thin metasurface with only an electric dipole response maximally absorbs 50 percent of the impinging light~\cite{Pozar2009,Tretyakov2014,Radi2015,Alaee2017}. Recently, metasurface perfect absorbers have been introduced using subwavelength resonant particles with \textit{electric and magnetic} responses~\cite{Engheta2002,Landy2008,Radi2015,Alaee2017}. These absorbers are thin compared to the operating wavelength and can fully absorb the light without a ground plane. An alternative approach to perfectly absorbing light in an optically thin metasurface with only an electric dipole response is coherent perfect absorber (CPA) by using the interference of two/multiple incoming beams~\cite{Chong:2010,Wan2011,Baranov2017}. In the CPA, it is essential to choose the intensities appropriately and the relative phases of the two input beams to achieve the perfect absorption. The position of the metasurface in the standing wave, produced by the interference of two coherent input beams, determines the amount of absorption.

%%%%%%%%%%%%%%%%%%%%%%%%%%%% FIGURE 1 %%%%%%%%%%%%%%%%%%%%%%%%%%%%%%%%%%%%%%%%%%%%%%%%%%%%%
\begin{figure}
    \centering
    \includegraphics[width=\linewidth]{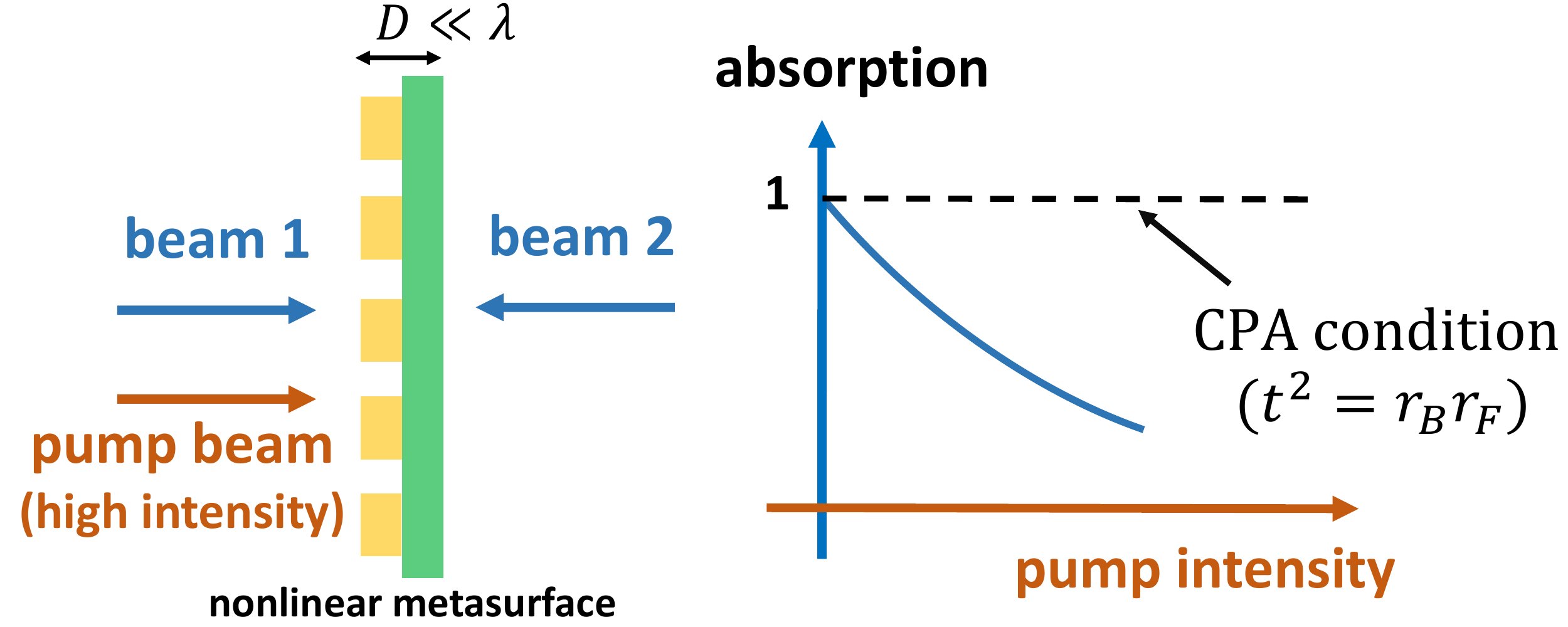}
    \caption{\textit{The Main idea of this letter:} Sketch of an ultrathin (i.e., $D \ll \lambda$, where D is the thickness of the metasurface) nonlinear metasurface illuminated by two counterpropagating incident beams. In the absence of a pump beam, the two coherent beams (i.e., beam1, beam2) lead to the coherent perfect absorption (CPA) when the reflected and transmitted lights destructively interfere on each side of the metasurface. CPA occurs when Fresnel coefficients satisfy $t^{2}=r_{\rm B}r_{\rm F}$. Note that single-beam illumination leads to partial absorption (see Fig.~\ref{fig:SingleBeam} (c) and Fig.~\ref{fig:SingleBeam} (d) ). In the presence of the pump, one can modify the absorption by increasing the pump intensity.}
 \label{fig:Main_idea}
\end{figure}
%%%%%%%%%%%%%%%%%%%%%%%%%%%%%%%%%%%%%%%%%%%%%%%%%%%%%%%%%%%%%%%%%%%%%%%%%%%%%%%%%%%%%%%%%%%%

Tunable perfect absorbers have been experimentally and theoretically realized by using tunable materials such as graphene~\cite{Alaee2012,Thongrattanasiri:2012}, liquid crystals~\cite{Shrekenhamer2013}, and phase-change materials (PCMs)~\cite{Kats:2012,Cao:2014,Alaee:2016}. Recently, a new class of low-index materials called epsilon-near-zero (ENZ) materials has attracted much attention because of its large nonlinear response~\cite{Alam:16,Alam2016nature,Caspani:2016,Reshef:19,Liberal:2017,Kinsey2019}. As the name indicates, ENZ materials are the materials in which the real part of the permittivity goes to zero at a particular frequency. 
ENZ materials exhibit a huge electric field enhancement as well as large nonlinearity at the ENZ frequency~\cite{kinsey:2019}.
In particular, indium tin oxide (ITO) and aluminum-doped
zinc oxide (AZO) as ENZ materials are used to achieve \textit{tunable} antennas and optical switches~\cite{Alam:16,Alam2016nature,Reshef:17,Reshef:19,Kinsey2019} and perfect absorbers~\cite{Bruno:2020}. Perfect absorption has been also achieved using the ENZ materials in the linear regime~\cite{Feng:2012,Halterman:2014,Zhong:2014,kim2016,Rensberg:2017}.

%% Feng with a ground plate, linear
%% Halterman:2014 Hyperbolic, linear
%% Zhong anisotropic, linear
%% Kim ITO CPA film
%% Rensberg film

In this letter, we propose an \textit{ultrathin nonlinear} coherent perfect absorber based on an ITO metasurface. In the absence of a pump beam, we show that the proposed metasurface exhibit coherent perfect absorption when illuminated from both sides. The CPA condition, i.e.  $t^{2}=r_{\rm B}r_{\rm F}$ occurs when one of the eigenvalues of the scattering-matrix is zero~\cite{Chong:2010}. We show that the proposed absorber is tunable when applying a third beam (pump beam) with high intensity~(see Fig.~\ref{fig:Main_idea}).  

%%%%%%%%%%%%%%%%%%%%%%%%%%%% FIGURE 2 %%%%%%%%%%%%%%%%%%%%%%%%%%%%%%%%%%%%%%%%%%%%%%%%%%%%%
\begin{figure}
    \centering
    \includegraphics[width=\linewidth]{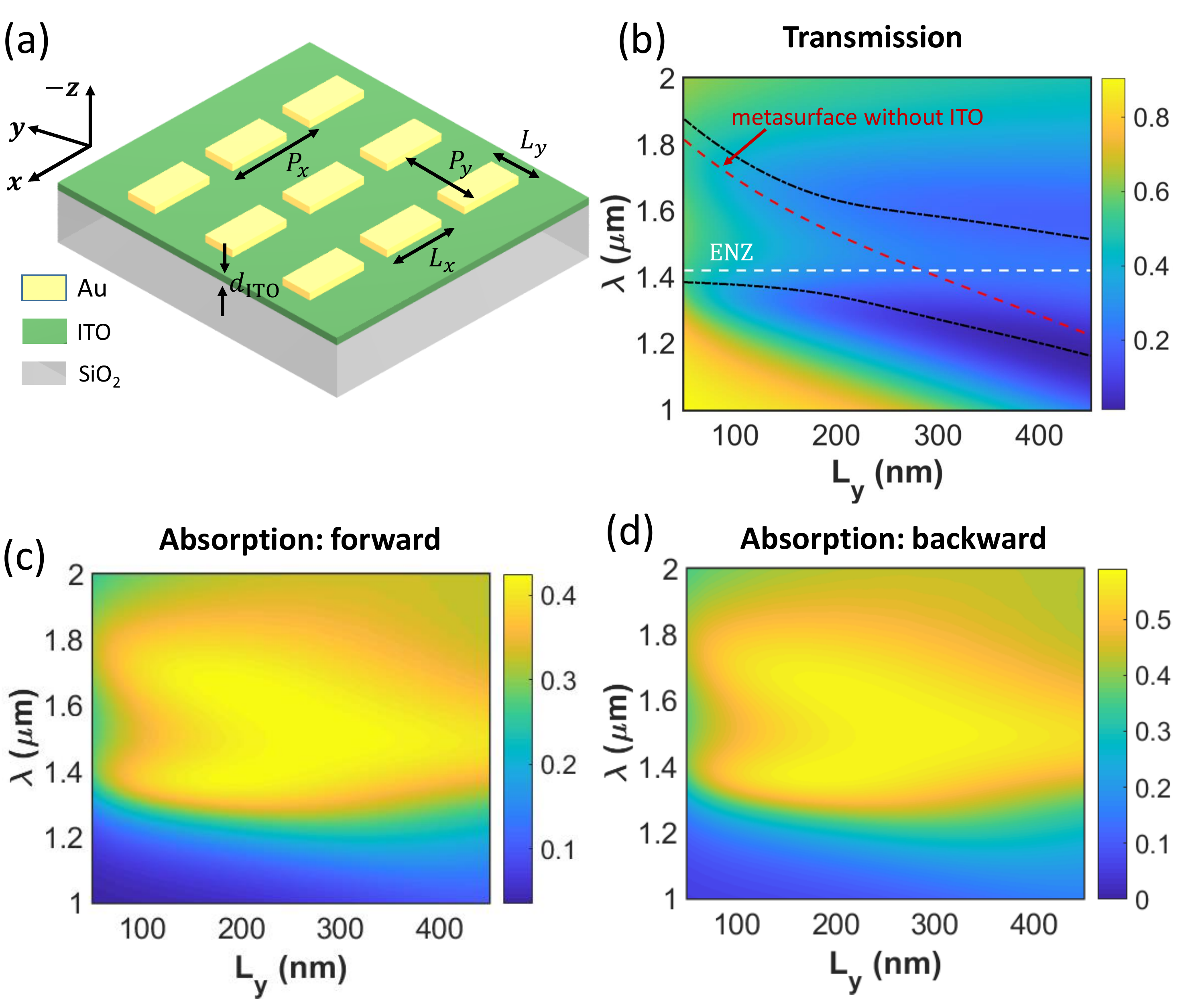}
    \caption{\textit{Single-beam excitation (forward and backward directions):} (a) Sketch of the tunable perfect absorber consisting of a gold metasurface on top of the indium tin oxide (ITO) coated $\rm SiO_{2}$. (b) Transmission as a function of wavelength and dimension of the gold antennas, i.e., $L_y$. The incident plane wave is polarized along the x-axis. Note that the dashed lines depict the resonance wavelength of the ENZ, i.e., $\lambda_{\rm ENZ}$~(white), and the resonance wavelength of the metasurface without ITO, i.e., $\lambda_{\rm MS}$~(red). Using Eq.~\ref{eq:Coupling}, the resonance wavelengths of the coupled system, i.e., $\lambda_{u,l}$ can be calculated (see the black dashed lines).  (c)-(d) Absorption ($A_{\mathrm{F/B}}=1-T-R_{\mathrm{F/B}}$) for forward ($\mathbf{k_{\rm F}}=k_0\mathbf{e}_z$) and backward ($\mathbf{k_{\rm B}}=-k_0n_{\rm SiO2}\mathbf{e}_z$) illuminations, respectively. The geometrical parameters of the metasurface are: $d_{\rm ITO}$ = 23\,nm, $P_x=P_y$ = 650\,nm, and $L_x$= 450\,nm. The thickness of the gold nanoantenna is $30$\,nm.} 
    \label{fig:SingleBeam}
\end{figure}
%%%%%%%%%%%%%%%%%%%%%%%%%%%%%%%%%%%%%%%%%%%%%%%%%%%%%%%%%%%%%%%%%%%%%%%%%%%%%%%%%%%%%%%%%%%%

\section{Theory and numerical results}
\textit{Linear response: single-beam excitation.—}The underlying physics of a coherent perfect absorber can be understood using the scattering-matrix-formalism~\cite{Haus1984,Yeh:2005}. The relation between the inputs $\mathbf{a}$ and outputs $\mathbf{b}$ waves can be described by $\mathbf{b}=\mathbf{S}\mathbf{a}$, where
\begin{equation}
\mathbf{a}=\left(\begin{array}{c}
a_{\rm F}\\
a_{\rm B}
\end{array}\right),\,\,\mathbf{b}=\left(\begin{array}{c}
b_{\rm F}\\
b_{\rm B}
\end{array}\right),\,\,\mathbf{S}=\left(\begin{array}{cc}
r_{\rm F} & t_{\rm B}\\
t_{\rm F} & r_{\rm B}
\end{array}\right).\label{eq:S_matrix}
\end{equation}
$r_{\rm F}$ and $r_{\rm B}$ are the reflection coefficients, $t_{\rm F}$ and $t_{\rm B}$ are the transmission coefficients from forward and backward directions, respectively~\cite{Yeh:2005}. Coherent perfect absorber occurs when the output waves vanishes, i.e., $\mathbf{b}=\boldsymbol{0}$  and that is $\mathbf{b}=\mathbf{S}\mathbf{a}_{\mathrm{CPA}}=\boldsymbol{0}$. This means that at least one eigenvalue of the $\mathbf{S}$ is zero, and the eigenvector is the CPA eigenmode, i.e., $\mathbf{a}_{\mathrm{CPA}}$. For symmetric structure, the eigenvectors are $\left(1,\pm1\right)$, i.e., the symmetric and antisymmetric input beams. Note that the CPA condition is the time reverse of the lasing~\cite{Chong:2010,Baranov2017}. 

Figure~\ref{fig:SingleBeam} (a) shows the schematic of the proposed tunable coherent perfect absorbers. The structure is composed of a gold metasurface of $30$\,nm thickness on a glass substrate. The ITO layer is sandwiched between the metasurface and the glass substrate. The permittivity of ITO is expressed as $\varepsilon=\varepsilon_{\mathrm{\infty}}-{\omega_{p}^{2}}/{\left({\omega}^2+i\gamma\omega\right)},$ where $\omega_p=2.67\times10^{15}\,\mathrm{rad/s}$, $\varepsilon_\mathrm{\infty}=3.9$, $\gamma=2.32\times10^{14}\,\mathrm{Hz}$~\cite{Alam:16}. The transmission spectrum is calculated in Fig.~\ref{fig:SingleBeam}~(b) for single-beam illumination (forward and backward directions) using a numerical finite element solver (COMSOL Multiphysics). Note that the forward and backward transmissions are identical, i.e., $ T=T_{\rm F}= T_{\rm B}$ because the system is reciprocal~\cite{Jalas:2013}. The structure exhibits strong coupling between two resonances, i.e. the bulk-plasma mode of the ENZ layer and the fundamental mode of the plasmonic nanoantennas~\cite{Alam2016nature,Bruno2020}. The resonances of the coupled system can be calculated from the eigenvalue problem, i.e.,  $H\psi=\hbar\widetilde{\omega}_{u,l}\psi$, 
where $\psi$ is the eigenstate of the coupled system and $H$ defined as~\cite{Agranovich:2003}
\begin{equation}
H=\hbar\left(\begin{array}{cc}
\widetilde{\omega}_{\mathrm{ITO}} & \Delta\\
\Delta & \widetilde{\omega}_{\mathrm{MS}}
\end{array}\right),
\end{equation}
where $\widetilde{\omega}_{\mathrm{ITO}}=\omega_{\mathrm{p}}-i\gamma$, $\widetilde{\omega}_{\mathrm{MS}}=\omega_{\mathrm{LSP}}-i\gamma_{\mathrm{LSP}}$. $\omega_{\mathrm{LSP}}$ and $\gamma_{\mathrm{LSP}}$ are resonance frequency and linewidth of the localized surface plasmon (LSP) mode of the plasmonic antennas in the absence of the ITO layer, respectively. $\Delta$ is the coupling between the bulk-plasma mode of the ITO and the fundamental mode of the metasurface. The eigenvalues of the coupled systems are given by
\begin{equation}
 \widetilde{\omega}_{u,l}=\frac{1}{2}\left(\widetilde{\omega}_{\mathrm{MS}}+\widetilde{\omega}_{\mathrm{ITO}}\pm\sqrt{\left(\widetilde{\omega}_{\mathrm{MS}}-\widetilde{\omega}_{\mathrm{ITO}}\right)^{2}+4\Delta^{2}}\right).\label{eq:Coupling}   
\end{equation}

The red and white dashed lines in Fig.~\ref{fig:SingleBeam}~(b) show the resonance of the metasurface without the ITO and the ENZ wavelength of ITO, respectively. Using Eq.~\ref{eq:Coupling}, we calculate the resonance wavelengths of the coupled system (see black dashed lines in Fig.~\ref{fig:SingleBeam}~(b)). There is a very good agreement between the simulation (using COMSOL) and the semi-analytical model (using Eq.~\ref{eq:Coupling}).

%%%%%%%%%%%%%%%%%%%%%%%%%%%% FIGURE 3 %%%%%%%%%%%%%%%%%%%%%%%%%%%%%%%%%%%%%%%%%%%%%%%%%%%%%
\begin{figure}
    \centering
    \includegraphics[width=\linewidth]{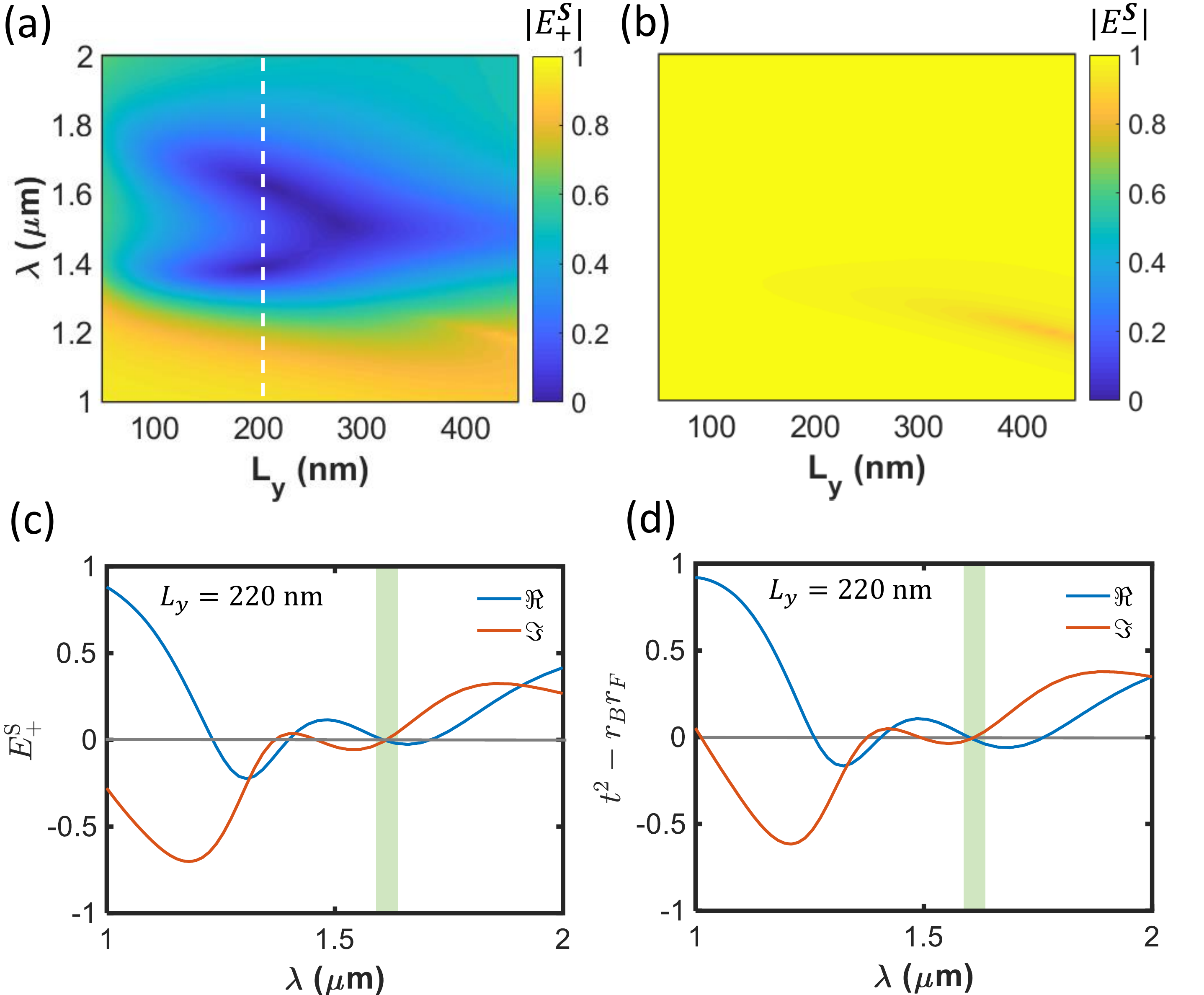}
    \caption{\textit{Coherent perfect absorber condition:} (a)-(b) Modulus of eigenvalues of the scattering matrix $S$ as a function of wavelength and dimension of the gold antennas. (c) Real and imaginary parts of the eigenvalue of the S-matrix as a function of wavelength for $L_y=220$\,nm. (d) $t^{2}-r_{\rm B}r_{\rm F}$ as a function of wavelength. The CPA condition occurs at 1.6$\mu\mathrm{m}$. Note that this condition is approximately satisfied  over a wide range of wavelengths.} 
    \label{fig:Eigenvalue}
\end{figure}
%%%%%%%%%%%%%%%%%%%%%%%%%%%%%%%%%%%%%%%%%%%%%%%%%%%%%%%%%%%%%%%%%%%%%%%%%%%%%%%%%%%%%%%%%%%%

Figure~\ref{fig:SingleBeam}~(c) and Fig.~\ref{fig:SingleBeam}~(d) show the absorption from the forward and backward directions, i.e., $A_{\mathrm{F/B}}=1-T-R_{\mathrm{F/B}}$. $R_{\rm F} = \left|r_{\rm F}\right|^{2}$ and $R_{\rm B} = \left|r_{\rm B}\right|^{2}$ are reflection from forward and backward, respectively. First, it can be seen that the absorption (or reflections) are \textit{not} symmetric for the two opposing irradiation directions. The asymmetric reflections/absorptions can be understood from the fact that the system is \textit{not} invariant under space inversion~\cite{Lindell1994,Alaee:2018PT1}. Note that, an optically thin metasurface consisting of an electric dipole antennas in a isotropic and homogeneous medium can only absorb 50 percent of the light ~\cite{Pozar2009,Radi2015,Alaee2017}. Our proposed asymmetric metasurface is ultrathin, i.e. $\lambda_{\rm res}/t_{\rm MS}\approx30$, where $\lambda_{\rm res}$ is the resonance wavelength and $t_{\rm MS}$ is the overall thickness of the metasurface (gold antennas and ITO). Following the  approach in Refs.~\cite{Pozar2009,Tretyakov2014,Alaee2017}, we found  that the maximum absorption for forward illumination is $A_{\rm F}^{\rm max}=n_{\rm air}/(n_{\rm air}+n_{\rm SiO_2})\approx0.4$~(see Fig.~\ref{fig:SingleBeam} (c)). Similarly, the maximum backward absorption read as $A_{\rm B}^{\rm max}=n_{\rm SiO_2}/(n_{\rm air}+n_{\rm SiO_2})\approx0.6$~(see Fig.~\ref{fig:SingleBeam} (d)). Interestingly, we get $A_{\rm B}^{\rm max}+A_{\rm F}^{\rm max} =1$.\\ 
To find the condition for the coherent perfect absorption, we calculate the two eigenvalues of the scattering-matrix, i.e., $\ensuremath{E_{\pm}^{\mathrm{S}}}=\left(-C_{\mathrm{FB}}\pm\sqrt{C_{\mathrm{FB}}^{2}-4\left(r_{\mathrm{F}}r_{\mathrm{B}}-t^{2}\right)}\right)/2$, where $C_{\mathrm{FB}}=-\left(r_{\mathrm{F}}+r_{\mathrm{B}}\right)$. Thus, the CPA condition, i.e. $t^2 = r_{\mathrm{F}}r_{\mathrm{B}}$ occurs when one of the eigenvalues of $\mathbf{S}$ matrix is zero, i.e., $E_{+}^{\rm S} = 0$. 
Figure~\ref{fig:Eigenvalue} (a)-(b) shows the modulus of the eigenvalues of $E_{\pm}^{\rm S}$ matrix as a function of frequency and length of the antenna $L_y$. It can be seen that only one of the eigenvalues, i.e., $E_{+}^{\mathrm{S}}$ vanish, e.g., at $L_y=220$\,nm (see the white dashed line in Fig.~\ref{fig:Eigenvalue} (a)). Figure~\ref{fig:Eigenvalue} (c) shows that the real and the imaginary parts of the eigenvalue $E_{+}^{\mathrm{S}}$ are zero at 1.6$\mu\mathrm{m}$ (see the green shadow). At this wavelength, Fresnel coefficients also satisfies $t^{2}=r_{\rm B}r_{\rm F}$ (see the green shadow in Fig.~\ref{fig:Eigenvalue} (d)). Note that the CPA condition is approximately satisfied over a wide range of wavelengths, and thus we expect to achieve a nearly broadband CPA for the two-beam illuminations. From now on, we will only consider the metasurface with $L_y=220$\,nm because this geometry satisfy the CPA condition, i.e., $t^{2}=r_{\rm B}r_{\rm F}$.

%%%%%%%%%%%%%%%%%%%%%%%%%%%% FIGURE 4 %%%%%%%%%%%%%%%%%%%%%%%%%%%%%%%%%%%%%%%%%%%%%%%%%%%%%
\begin{figure}
    \centering
    \includegraphics[width=\linewidth]{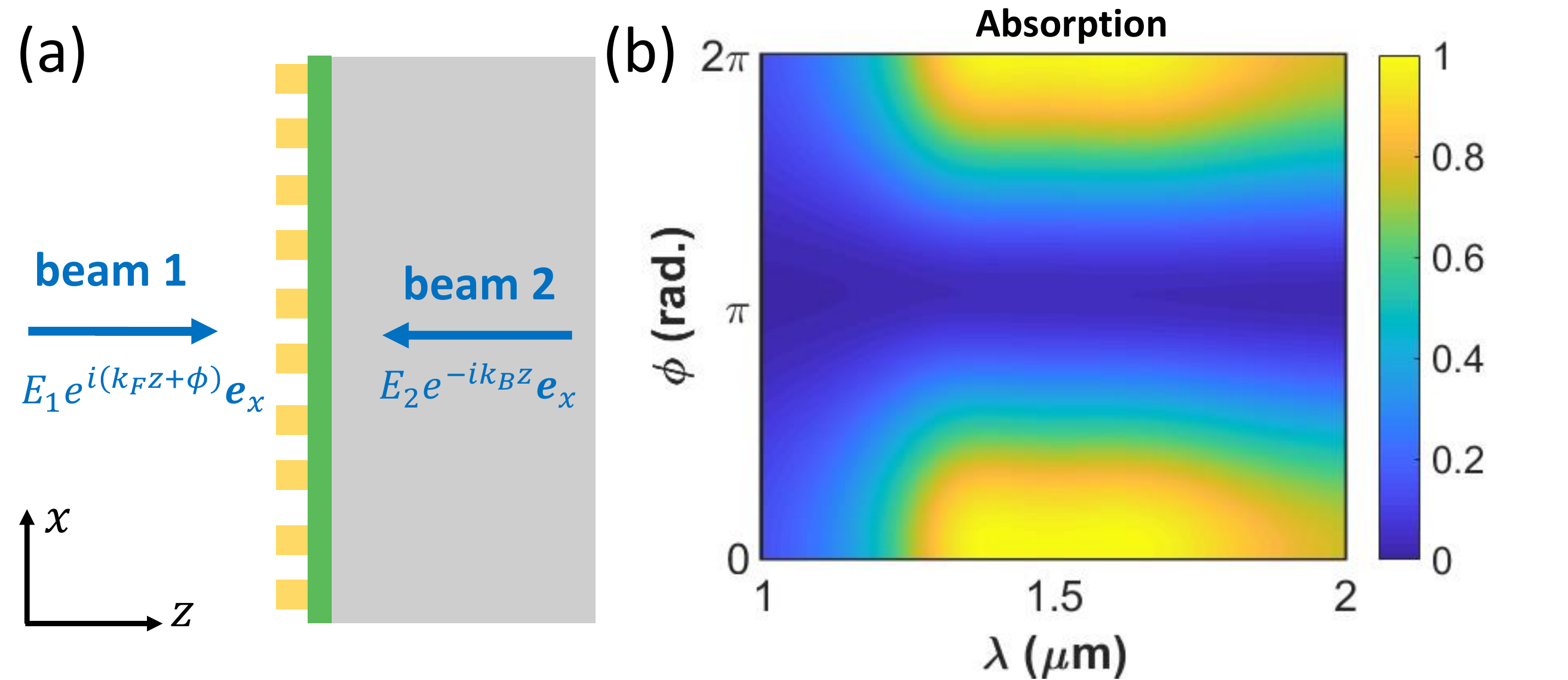}
    \caption{\textit{Two counterpropagating incident beams:} (a) Schematic representation of the CPA metasurface illuminated by two input beams with low intensities. (b) Absorption as a function of the wavelength and the relative phase between input beams, i.e., $\phi$. The CPA occurs when two beams are in phase, i.e., $\phi=0, 2\pi$. The absorption is nearly zero when two input beams are out of phase, i.e., $\phi=\pi$.}  \label{fig:TwoBeam}
\end{figure}

%%%%%%%%%%%%%%%%%%%%%%%%%%%%%%%%%%%%%%%%%%%%%%%%%%%%%%%%%%%%%%%%%%%%%%%%%%%%%%%%%%%%%%%%%%%%

\textit{Linear response: two counterpropagating incident beams.—}Let us consider the CPA configuration (see the two-beam illumination in Fig.~\ref{fig:TwoBeam} (a)). For the two beams, the joint absorption can be defined as~\cite{Baldacci:2015}
\begin{equation}
A=1-\frac{\left|b_{1}\right|^{2}+\left|b_{2}\right|^{2}}{\left|a_{1}\right|^{2}+\left|a_{2}\right|^{2}}. \label{Eq:JointAbsorption}
\end{equation}
For lossless systems, absorption is zero $A = 0$ and unity $A = 1$  for the CPA. Figure ~\ref{fig:TwoBeam} (b) shows the joint absorption for the two-beam illumination as a function of the relative phase of the input beams, i.e., $\phi$. By varying the relative phase of the two-beam from 0 to $\pi$, the absorption of the  metasurface goes from perfect absorption to nearly zero absorption while keeping the two beam intensities constant. Note that we used different amplitudes for the incident beams, i.e., $E_2 = 0.86E_1$, which is related to CPA eigenvector of the $\mathbf{a}_{\rm CPA}$. In other words, the Fresnel coefficients are different because the system is not symmetric. Nearly perfect absorption occurs over a large range of frequencies, i.e., $1.3-1.75\,\mu\mathrm{m}$ (see also the white dashed line in Fig.~\ref{fig:Eigenvalue} (a), which depicts the CPA condition). Figure ~\ref{fig:TwoBeam} (b) shows that when two beams are out of phase, the absorption is nearly zero because of the vanishing incidence field inside the metasurface.    

\begin{figure}
    \centering
    \includegraphics[width=\linewidth]{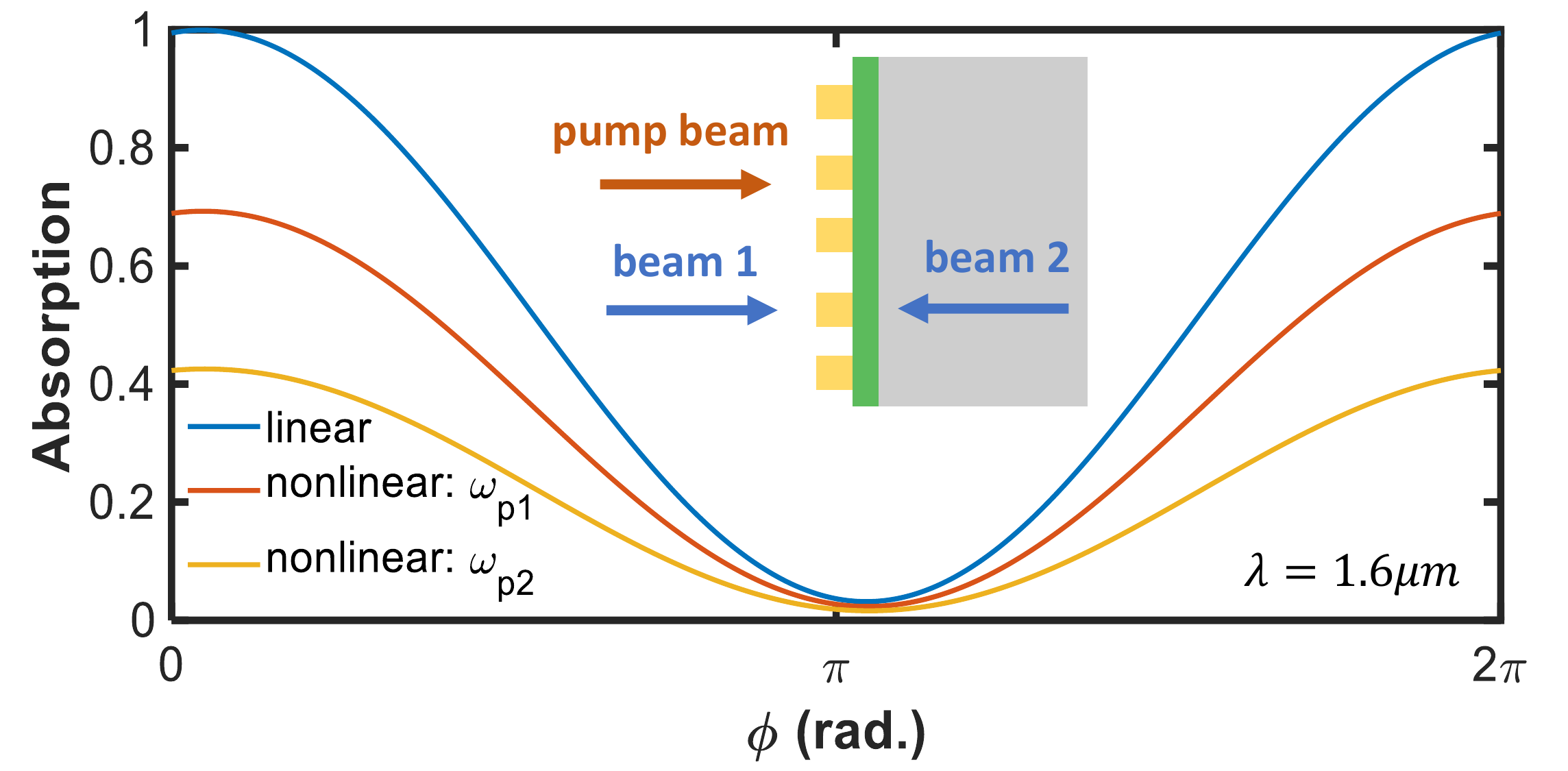}
    \caption{\textit{Two counterpropagating input beams with a pump beam:} (a) Absorption as a function of the relative phase between two beams at wavelength $1.6\,\mu\mathrm{m}$. The inset shows the sketch of a tunable coherent perfect absorber illuminated by two low-intensity counterpropagating input beams and a high-intensity pump beam. The pump beam changes the refractive index of the ITO. In the nonlinear simulation, we assumed that $\omega_{p1} = 2\times10^{15}\,\mathrm{rad/s}$ and $\omega_{p2} =1.33\times10^{15}\,\mathrm{rad/s}$. Considering the field enhancement in the ITO layer, these plasma frequencies approximately corresponds to pump intensities $I_0=5-10\,\rm GW/cm^2$.} \label{fig:Three beam}
\end{figure}

\textit{Nonlinear response.—}Next, we consider the effect of a pump beam on the absorption of the proposed nonlinear metasurface. It has been experimentally shown that the refractive index of the ITO depends on the intensity pump beam~\cite{Alam:16,Reshef:17}. At the high optical intensity regime, ITO's response can be described by the Drude model~\cite{Alam:16,Alam2016nature,Wang:2019}. In this model, the plasma frequency of the ITO decreases as the intensity of light increases~\cite{Alam:16,Alam2016nature,Wang:2019}. In our simulation, we changed the plasma frequency in the Drude model to model ITO's nonlinearities. Figure ~\ref{fig:Three beam} shows the absorption as a function of the relative phase between two beams, i.e., $\phi$ with and without the pump beam. It can be seen that the absorption is significantly decreased by applying the pump beam (from unity to 0.4) when the relative phase of the two-beam is zero.
%%%%%%%%%%%%%%%%%%%%%%%%%%%% FIGURE 5 %%%%%%%%%%%%%%%%%%%%%%%%%%%%%%%%%%%%%%%%%%%%%%%%%%%%%

In conclusion, we have shown that coherent perfect absorption can occur for a metasurface made of optical nanoantennas coupled to an epsilon-near-zero material. In particular, we have demonstrated that the absorption of the metasurface can be tuned by applying a high-intensity pump beam. Our results can be employed to design tunable thermal emitters, optical modulators, optical sensors, and optical data process. 

\paragraph{Acknowledgment and Funding} R. A. is grateful to  M. Karimi, M Z. Alam, O. Reshef, B. Braverman, and J. Upham for helpful discussions and acknowledges the support of the Alexander von Humboldt Foundation through the Feodor Lynen Fellowship. The authors acknowledge support through the Natural Sciences and Engineering Research Council of Canada, the Canada Research Chairs program, and the Canada First Research Excellence Fund. 

% Bibliography

\end{document}